\documentstyle[preprint,aps]{revtex}

\begin{document}
\draft
\preprint{\small gr-qc/9306024 Alberta-Thy-30-93 May '93}

\title
{Faraday lines and observables for the Einstein-Maxwell theory}

\author{Viqar Husain}

\address
 {Theoretical Physics Institute,
 University of  Alberta,\\
Edmonton, Alberta, Canada T6G 2J1}
\maketitle

\begin{abstract}

In recent work on Einstein gravity in four dimensions using the
Ashtekar
 variables, non-local loop variables have played an important role in
 attempts to formulate a quantum theory. The introduction of such
variables
 is guided by gauge invariance, and here an infinite set of loop
variables
is introduced for the Hamiltonian form of the Einstein-Maxwell
theory.
The loops that enter the description naturally are the (source free)
 electric field lines.
These variables  are invariant under spatial diffeomorphisms and they
also form a closed Poisson algebra. As such they may be useful for
quantization attempts and for studying classical solutions.

\end{abstract}
\pacs{PACS numbers: 4.20.-q, 4.20.Fy, 4.20.Me, 4.60.+n}
\vfill
\eject

The canonical approach to quantum gravity in four dimensions
has received much attention in recent years due to the introduction
by Ashtekar of new Hamiltonian variables for general relativity
\cite{ash}.
These phase space variables, an electric field and vector potential,
are
similar to those for SO(3) Yang-Mills theory,
and so allow the introduction of holonomies or Wilson loops as
natural
variables on the phase space.

Classical phase space variables based on loops have been of
substantial
 use in canonical quantum gravity.
Rovelli and Smolin introduced an infinite
set of gauge invariant loop variables  on the Ashtekar phase space
which are made from holonomies of the vector potential traced
together with
the conjugate electric field.
These variables form a closed Poisson algebra, and a representation
of this algebra on functions of loops has been used to study the
quantum theory \cite{lc}. The Hamiltonian and spatial diffeomorphism
constraints  of general relativity may be written  as operators on
this
loop space  \cite{lc,blenbrugpull}.
The main result of this work is that there are an infinite
set of solutions of the Dirac quantization conditions in the loop
representation, and the physical states so obtained are labelled by
knot invariants \cite{lc}.
This labelling arises essentially because the diffeomorphism
invariant
  information in a loop is its knot class.

 Many open problems remain in this approach, and among them is the
question of what classical variables are to be represented as
Hermitian
operators on the space of physical states. The natural answer to this
is that the variables should be invariant under all the gauge
symmetries
of the theory. However requiring this for general relativity amounts
to
finding constants of the motion, since the Hamiltonian constraint
generates
both gauge and dynamics. In particular, a complete set of fully gauge
 invariant variables amounts to complete integrability of the full
 Einstein equations. More practical may be a weaker condition,
 namely requiring gauge  invariance under only the kinematical gauge
 symmetries - spatial diffeomorphisms and Yang-Mills symmetries. The
  classical and quantum evolution of these variables via the
Hamiltonian
constraint can then be studied.

 The Rovelli-Smolin loop variables mentioned above are not invariant
 under spatial diffeomorphisms. In pure gravity it isn't possible
 to construct loop variables that are spatially diffeomorphism
invariant,
essentially because loops are external to the phase space, and phase
space
functionals that depend on external variables cannot be made
diffeomorphism
 invariant.
 On the other hand loop variables seem essential to constructing a
loop
 representation, and it is only in this representation that a large
number
 of solutions of the Dirac quantization conditions have been
obtained. It
 is therefore natural to ask if there is a way of obtaining
diffeomorphism
 invariant loop observables when there is coupling to matter fields.
This
 is important also from the point of view of incorporating matter
into
 the loop space formalism, and the idea presented below may provide a
way
  of doing this.

 In this letter the question of gauge invariant variables is
discussed when
 there is coupling of Einstein gravity to the Maxwell field. First
the
 Einstein-Maxwell Hamiltonian system is given in the Ashtekar
variables
 and the Maxwell Gauss law is solved classically to obtain a
 partially reduced system. Phase space loop variables that are
invariant
 under the gravity Gauss law and spatial diffeomorphism constraints
 are then presented. This is the main result of the paper. This
 is followed by a discussion of the possible uses of these variables.

 The phase space variables for the Einstein-Maxwell system may be
taken as
  the Ashtekar variables $(A_a^i, E^{ai})$ for the gravitational
field, and
 the pairs $(a_a, e^a)$ for the Maxwell field. $a,b,c...$ are spatial
  indices and $i,j,...$ are the internal SO(3) indices. The phase
   space constraints reflecting the symmetries are
 \begin{eqnarray}
 & & \partial_ae^a = 0, \\
 & &  D_a E^{ai} = 0, \\
 & &  F_{ab}^iE^{bi} + f_{ab}e^b = 0, \\
 & &  \epsilon^{ijk}F_{ab}^k E^{ai} E^{bj} +
 ({\rm det} E)^{-2}E^{ai}E^{ci}E^{bj}E^{dj}\bigl( e_{ab}e_{cd} +
f_{ab}f_{cd}
 \bigr) = 0.
 \end{eqnarray}
 where $D_a$ is the covariant derivative of $A_a^i$ and $F_{ab}^i$ is
its
 curvature, det$E$ is the determinant of the inverse of $E^{ai}$,
 $e_{ab}=\epsilon_{abc}e^c$, and $f_{ab}=\partial_{[a} a_{b]}$. The
first two
 constraints are the Maxwell and gravity Gauss law constraints, the
third
 is the spatial diffeomorphism constraint, and the last is the
Hamiltonian constraint. The total Hamiltonian $H$ is a linear
combination of these constraints and the phase
space variables evolve via the Hamilton equations.

Starting with this constrained system, the goal is to first extract
at the
classical level the two local unconstrained degrees of freedom of the
Maxwell field, and then eliminate $(a_a,e^a)$ in favor of these in
the
constraints. This partially reduced system of constraints will then
be used
 as the starting point for finding the gauge invariant observables.

 The reduction may be done by
finding a solution of the  Hamilton-Jacobi equation associated with
the Maxwell Gauss law constraint \cite{ct}, which is obtained by the
replacement $e^a\rightarrow \delta S/\delta a_a$ in (1). The solution
must
have two  integration variables $u_n$ ($n=1,2$) corresponding to the
two
 degrees of freedom. A solution is
\begin{equation}
S[a;u_1,u_2]=\int d^3x\ \epsilon^{abc}a_a\partial_bu_1\partial_cu_2
\end{equation}
The electric field $e^a$ is then given by
\begin{equation}
e^a = {\delta S\over \delta a_a} =
\epsilon^{abc}\partial_bu_1\partial_cu_2,
\end{equation}
which solves the Maxwell Gauss law.
The momenta conjugate to the reduced variables $u_n$ are given in the
usual
way from the Hamilton-Jacobi functional:
\begin{eqnarray}
p_{u_1} & = &{\delta S \over \delta u_1} =
{1\over 2}\epsilon^{abc}f_{ab}\partial_c u_2 \\
p_{u_2} & = &{\delta S \over \delta u_2} =
-{1\over 2}\epsilon^{abc}f_{ab}\partial_c u_1
\end{eqnarray}

These relations may be used to eliminate the six local $(a_a,e^a)$
variables
 for the four reduced ones $(u_n,p_{u_n})$ in the constraints (3) and
(4).
 This  gives the reduced system.
 In particular the diffeomorphism constraint (3) reduces to
 \begin{equation}
 C_a \equiv F_{ab}^iE^{bi} + p_{u_n}\partial_au_n = 0
 \end{equation}

  The aim now is to find a set of observables $T$ that are invariant
 under the transformations generated by the gravity Gauss law (2) and
the
 reduced spatial diffeomorphism constraints (9). By definition, the
$T$
 must have weakly vanishing Poisson brackets with these kinematical
 constraints.

 The configurations
 of the Maxwell fields $u_1=c_1$ and $u_2=c_2$ for constants
$c_1,c_2$
 define two surfaces, and their intersection gives a loop. Since
 $e^a=\epsilon^{abc}\partial_b u_1\partial_cu_2$, the electric field
lines are in fact parallel to these loops. The idea now is to
 define the gravity loop observables based on these electric field
loops
 $\gamma[u_n](c_1,c_2)$  rather than auxiliary loops. Such a
construction
 will, if done properly, make all the gravity loop variables
\cite{lc}
 invariant under spatial diffeomorphisms as well.

 The first loop observable is the trace of the holonomy, and with the
loops
 chosen as above it becomes diffeomorphism invariant:
 \begin{equation}
 T^0[A,u_n](c_1,c_2)={\rm Tr}U\equiv{\rm Tr Pexp}\int_{\gamma[u_n]}
dx^a A_a
 \end{equation}
 The next two observables have one and two insertions of the triad
$E^{ai}$
  in the holonomies and are defined by
\begin{eqnarray}
T^1[A,E,u_n](c_1,c_2)& = &\int_{\gamma[u_n]} ds\ w_a(s)
    Tr[ E^a(\gamma(s)) U_\gamma(s,s)] \\
  T^2[A,E,u_n](c_1,c_2)& = &\int_{\gamma[u_n]} ds\
\int_{\gamma[u_n]} dt\
w_a(s) w_b(t) \nonumber \\
&  &\times  Tr[ E^a(\gamma(s
))U_\gamma(s,t)E^b(\gamma(t))U_\gamma(t,s)],
\end{eqnarray}
where the 1-form  density
\begin{equation}
w_a\equiv \epsilon_{abc}
 \dot{\gamma}^b {\delta\gamma^c\over \delta u_1}.
 \end{equation}
$\dot{\gamma}^a =dx^a/ds$ is the tangent vector to the loop, $s$ is
 a parameter along the loop, and $U(s,t)$ is the holonomy between the
 parameter values $s,t$.
  In (13),  $u_1$ may be replaced by $u_2$ to give distinct
observables.
For comparison,  the traces in the integrands of (11) and (12) are
the first two Rovelli-Smolin loop variables \cite{lc}.

The generalizations $T^N$ of these for $N$ triad insertions are
obtained in a
similar way. These observables are functionals of $A_a^i,E^{ai},u_n$,
and functions of the two parameters $c_1,c_2$, and are independent
of the momenta $p_{u_n}$ conjugate to the $u_n$.
The Poisson brackets $\{T^M,T^N\}$  are therefore determined solely
by the
gravitational  variables, and are non-zero for any two $T^N$
only if the corresponding loops intersect. More generally, the loops
may be determined by $f(u_1,u_2)=c_1$ and $g(u_1,u_2)=c_2$ for
arbitrary
functions $f,g$. So given a configuration of the phase space
variables
and two arbitrary functions $f,g$, the loop variables will in general
not commute and the Poisson algebra  will be of the form
\begin{equation}
\{T^M,T^N\} \sim T^{M+N-1}
\end{equation}
This is of the same form as that of the gravity loop variables with
auxiliary
loops \cite{lc}.

Since the $T^N$  described above do not depend on all the phase space
variables, they cannot form a complete set of observables.
Observables
depending on the momenta are $P_n=\int d^3x\ p_{u_n}$ but with these,
the Poisson algebra no longer retains the nice form (14). An infinite
number of new observables are generated from the $T^N$  by Poisson
brackets
with $P_n$, since these act to shift the loops and introduce
additional
 derivatives on the 1-form densities
 $w_a$. The algebra can of course be completed by exhausting all
possible
 Poisson brackets.

 For the quantum theory, the Dirac quantization conditions
  in the configuration representation are obtained by the
replacements
 $E^{ai}\rightarrow \delta/\delta A_a^i$ and $p_{u_n}\rightarrow
 \delta/\delta u_n$ in the constraints. Then the  wavefunctional
 \begin{equation}
 \Psi[A,f(u_1,u_2),g(u_1,u_2)](c_1,c_2)={\rm TrPexp}
 \int_{\gamma[f,g]}dx^a\ A_a
 \end{equation}
solves the diffeomorphism and Gauss law conditions. This solution
is parametrized by two arbitrary functions $f$ and $g$, and the
constants
 $c_1,c_2$. Furthermore
it is also annihilated by the gravitational part of the Hamiltonian
constraint. The full action of the Hamiltonian constraint (4) is
complicated
 by the determinant terms in the Maxwell part. This  latter term
 is in fact  non-local when the inversions of (7) and (8) are
substituted
   in (4).

For the quantum theory in the loop representation,  the $T^N$,
together
with all the observables required to complete them, provide a natural
set of observables to represent as operators on the solution space
of the constraints. These must commute with the diffeomorphism
operator
 and have the appropriate evolution with respect to the Hamiltonian
 constraint operator.

The basic idea that is utilized above to construct diffeomorphism
invariant
 observables is the use of matter fields to define loops. This is
related
 to the general idea of  using matter reference systems for
gravitational
variables, and it has a long history.  It is discussed for a point
particle  by DeWitt \cite{dw}, and more recent discussions in related
contexts have been by Rovelli \cite{c}, and by Kuchar and Torre
\cite{kt}.
The latter have used particular types of matter which is defined by
 certain coordinate conditions. Their approach fixes all the
coordinates
 by reference to matter fields, whereas here only the loops have been
  fixed by matter configurations. A similar approach to loop
variables for
  couplings to scalar fields has also been discussed by the
author\cite{v}.
 In the context of the Ashtekar variables, matter fields have been
 used to construct surface observables using scalar fields
\cite{csurf},
  and antisymmetric tensor fields \cite{lsurf}. These surface
observables
are well defined in terms of loop observables, and they
exibit a discrete spectrum in the loop representation, with the areas
quantized in units of the Planck area \cite{alc}.

In summary,  an infinite set of diffeomorphism invariant loop
observables

for the Einstein-Maxwell system have been described. The construction
used

electric field lines as the reference systems for the gravitational
loop

 variables.  It is clearly possible

 to construct loop observables of this kind using any

 matter fields.


\begin{references}

\bibitem{ash} A. Ashtekar, {\it Non-perturbative canonical gravity},
Lecture notes in collaboration with R. S. Tate

(World Scientific, Singapore, 1991).

\bibitem{lc} C. Rovelli and L. Smolin, Phys. Rev. Lett. {\bf 61},
1155

(1988); Nucl. Phys. {\bf B331}, 80 (1990).

\bibitem{blenbrugpull} M. Blencowe, Nucl. Phys. {\bf B341} 213
(1990);
 B. Brugmann and J. Pullin, Nucl. Phys. {\bf B390} 399 (1993).

\bibitem{ct} E. T. Newman and C. Rovelli, Phys. Rev. Lett. {\bf 69},
1300

(1992).

\bibitem{dw} B. S. DeWitt, in {\it Gravitation, an intorduction to
current
research}, edited by . L. Witten  (Wiley, New York, 1962).

\bibitem{c} C. Rovelli, Class. and Quant. Gravity {\bf 8} 297 (1991).

\bibitem{kt} K. V. Kuchar and C. G. Torre,  Phys. Rev. D {\bf 43} 419
(1991);

Phys. Rev. D {\bf 44} 3116 (1991).


\bibitem{v} V. Husain, {\it General covariance, loops, and matter},
gr-qc 9304010, to appear in Phys. Rev. D. (1993).

\bibitem{csurf} C. Rovelli, {\it A generally covariant quantum field
theory},
University  of Pittsburgh preprint (1992).

 \bibitem{lsurf} L. Smolin, {\it Diffeomorphism invariant observables
  in quantum gravity from a dynamical theory of surfaces}, Syracuse
University preprint (1992);
 {\it Time, measurement and information loss in quantum cosmology},
in
  {\it Directions in General Relativity} Vol 1, edited by. B.L. Hu
and
  T. A. Jacobson, (Cambridge University Press, Cambridge 1992).

\bibitem{alc} A. Ashtekar, C. Rovelli, and L. Smolin, Phys. Rev.
Lett.

 {\bf 69}, 237 (1992).

\end{references}
\end{document}